\documentclass{aa}
\usepackage[T1]{fontenc}
\usepackage{natbib}
\usepackage{csquotes}
\usepackage{xspace}
\usepackage{txfonts}
\usepackage{amsmath}
\usepackage{threeparttable}
\usepackage{graphicx}
\usepackage{hyperref}
\usepackage[separate-uncertainty=true]{siunitx}
\usepackage{tikz}
\usepackage{subcaption}
\usepackage{placeins}
\usepackage{algorithm}
\usepackage{algpseudocode}

\usepackage{color}

\begin{document}

\title{Harnessing machine learning for accurate treatment of overlapping opacity species in general circulation models}
\titlerunning{Harnessing machine learning for accurate treatment of overlapping opacity species in general circulation models}
\authorrunning{Schneider et al.}
\author{Aaron David~Schneider$^{1,2}$, Paul Molli{\`e}re$^{3}$, Gilles Louppe$^{4}$, Ludmila Carone$^{5}$, Uffe Gr\r{a}e~J{\o}rgensen$^{1}$, Leen Decin$^{2}$ \& Christiane Helling$^{5}$}
\institute{
  (1) Centre for ExoLife Sciences, Niels Bohr Institute, {\O}ster Voldgade 5, 1350 Copenhagen, Denmark\\
  (2) Institute of Astronomy, KU Leuven, Celestijnenlaan 200D, 3001, Leuven, Belgium\\
  (3) Max-Planck-Institut für Astronomie, K{\"o}nigstuhl 17, 69117 Heidelberg, Germany\\  
  (4) Montefiore Institute, University of Li{\`e}ge, Li{\`e}ge, Belgium\\   
  (5) Space Research Institute, Austrian Academy of Sciences, Schmiedlstrasse 6, A-8042 Graz, Austria\\    
} \date{\today}
\offprints{A. Schneider,\\ \email{aaron.schneider@nbi.ku.dk}\\or \email{aarondavid.schneider@kuleuven.be}}
\abstract{To understand high precision observations of exoplanets and brown dwarfs, we need detailed and complex general circulation models (GCMs) that incorporate hydrodynamics, chemistry, and radiation. For this study, we specifically examined the coupling between chemistry and radiation in GCMs and compared different methods for the mixing of opacities of different chemical species in the correlated-k assumption, when equilibrium chemistry cannot be assumed. We propose a fast machine learning method based on DeepSets (DS), which effectively combines individual correlated-k opacities (k-tables). We evaluated the DS method alongside other published methods such as adaptive equivalent extinction (AEE) and random overlap with rebinning and resorting (RORR). We integrated these mixing methods into our GCM (expeRT/MITgcm) and assessed their accuracy and performance for the example of the hot Jupiter HD~209458 b. Our findings indicate that the DS method is both accurate and efficient for GCM usage, whereas RORR is too slow. Additionally, we observed that the accuracy of AEE depends on its specific implementation and may introduce numerical issues in achieving radiative transfer solution convergence. We then applied the DS mixing method in a simplified chemical disequilibrium situation, where we modeled the rainout of TiO and VO, and confirmed that the rainout of TiO and VO would hinder the formation of a stratosphere. To further expedite the development of consistent disequilibrium chemistry calculations in GCMs, we provide documentation and code for coupling the DS mixing method with correlated-k radiative transfer solvers. The DS method has been extensively tested to be accurate enough for GCMs; however, other methods might be needed for accelerating atmospheric retrievals.
}


\maketitle
\section{Introduction}\label{sec:introduction}
General circulation models (GCMs) have been applied with a lot of success to understand the 3D nature of exoplanetary atmospheres \citep[for a review see][]{Showman2020review}. These models usually consist of a dynamical core that solves the equations of hydrodynamics, coupled to physical parameterizations with differing complexity that describe the forcing on the temperature and velocity field. A very common physical parameterization in GCMs is that of heating and cooling by (gas) radiative transfer.

With the advent of detailed spectra from medium-resolution space-based telescopes such as JWST, we will soon have the ability to map the spatial distribution of chemical species in the atmospheres of hot gas giants. Ground-based high-resolution spectroscopy already allows us to measure chemical variations between morning and evening terminators \citep[e.g.,][]{Ehrenreich2020W76,Kesseli2021W76,Kesseli2022W76}. To comprehend these spatial distribution maps, we need 3D numerical models that can couple hydrodynamics, chemistry, and radiative transport. Only one GCM, the \texttt{unified model} (UM), currently possesses this capability \citep{Drummond2020GCMChem,Zamyatina2023GCMchem}. \citet{Lee2023GCMChem} employed a faster chemical network, but did not incorporate a chemically consistent radiative transfer solver. However, these studies consistently indicate long runtimes. 

One of the challenges of such models is the consistent coupling between chemical abundances and line opacities as used in the radiative transfer \citep{Amundsen2017ck}. Line opacities of molecular species in the pressure-temperature range of warm and hot gas giants often are a collection of millions of lines, which must also be accurately accounted for in low resolution (fast) radiative transfer. It was therefore realized early on that rapid calculations would need some simplifications. One of these is the correlated-k method \citep{Goody1989c-k}, which is similar to the method of opacity distribution functions (ODFs) introduced for stellar atmospheres \citep{Gustafsson1975c-k}. The correlated-k method as well as the ODF method converts the wavelength-dependent opacity into distribution functions (called k-tables or ODFs) by sorting opacity values within spectral bins (see Sect.~\ref{sec:mixing_intro}). This approach captures the dynamic range of the opacity and allows radiative transfer calculations with a small set of spectral bins (usually five to 30 for GCMs and a few hundred for spectra) and an accuracy of a few percent on bolometric fluxes \citep[e.g.,][]{Amundsen2014GCM,Leconte2021exok,Schneider2022GCM}.

The main drawback of the correlated-k method is the loss of all wavelength information in a spectral bin, when the opacity is converted to k-tables. Since the wavelength distribution of the opacity is depth-dependent, the loss of wavelength information at first-order approximation causes erroneous optical depth calculation through the atmosphere, and hence erroneous radiative transfer computation and energy balance. This can be corrected for in statistical ways, as described in detail below using the RORR method (see Sect.~\ref{sec:RORR}). In the corresponding ODF scheme traditionally used in older stellar atmosphere computations, the problem is the same and was discussed for example in \citet{Saxner1984ODF}, where it was concluded that the cost in increased computing time as function of the number of individual opacity species made it unfeasible to continue the ODF scheme for cooler stars. Newer stellar models are therefore usually computed based on the opacity sampling scheme, as discussed for example in \citet{Jorgensen1992samplingRT}, \citet{Gustafsson1994}, and \citet{Helling1998RTOpac}. It is, however, not obvious how one should treat the strong atomic lines correctly in the opacity sampling scheme, and if these are important in exoplanetary atmospheres alongside with the multitude of molecular lines, the correlated-k method may turn out to be superior, or a new hybrid method may be needed. 

In order to still use the correlated-k method, GCMs often use premixed k-tables, which tabulate the k-tables for a given chemical mixture assuming that the gas is in chemical equilibrium and abundances can be constrained as a function of pressure and temperature alone \citep[e.g.,][]{Showman2009GCM,Lee2021GCM,Schneider2022GCM,Deitrick2022GCM}. The k-table at a certain grid point is then recovered by interpolating on the pressure-temperature dependent premixed grid of k-tables. This approach, however, is not chemically correct when the gas is not in chemical equilibrium, which is the case if processes such as chemical kinetics or photochemistry are taken into account in the model.

The UM \citep{Amundsen2016UKMetGCM} is, to our knowledge, currently the only hot Jupiter GCM that can handle k-table mixing during runtime, without the need of premixed tables. Using a 1D radiative transfer code, \citet{Amundsen2017ck} benchmarked several numerical schemes that can approximate k-table mixing. Treating the k-table mixing during runtime rather than using premixed tables, introduces more freedom in the radiative transfer computation, but for gasses with many opacity sources only if it can be performed sufficiently fast. We therefore extend upon \citet{Amundsen2017ck} in this paper, by introducing a new machine learning accelerated technique, and by coupling these methods to our GCM (\texttt{expeRT/MITgcm}). This paper starts by introducing the correlated-k method and several approximate methods for k-table mixing in Sect.~\ref{sec:mixing_intro}. We then discuss the setup of our GCM in Sect.~\ref{sec:methods}, show the benchmarking results in Sect.~\ref{sec:results}, and a simple disequilibrium application in Sect.~\ref{sec:rain}, where we apply our model in a simple rainout situation, where condensation of TiO and VO are approximated. We finally conclude and discuss the implications of this work in Sect.~\ref{sec:discussion}.


\section{Mixing species}\label{sec:mixing_intro}
According to the Lambert Beer law, the total transmission $\mathcal{T}$ of light passing through a homogeneous slab of gas with density $\rho$ [\SI{}{\kg\per\cubic\m}] opacity $\kappa$ [\SI{}{\square\m\per\kg}] and thickness $d$ [\SI{}{\m}] in the spectral window from frequency $\nu_0$ [\SI{}{\Hz}] to $\nu_1$ [\SI{}{\Hz}] is given by
\begin{equation}\label{eq:transm}
	\mathcal{T} = \frac{1}{\nu_1-\nu_0}\int_{\nu_0}^{\nu_1} \exp\left(-\kappa(\nu)\rho d\right)\mathrm{d}\nu.
\end{equation}
In the correlated-k method, the integral of Eq.~\ref{eq:transm} is solved by substituting $\nu$,
\begin{equation}\label{eq:transm_g}
	\mathcal{T} = \frac{1}{\nu_1-\nu_0}\int_{0}^{1} \exp\left(-\kappa(g)\rho d\right) \mathrm{d}g,
\end{equation}
such that the integration over frequency is substituted by an integration over a new independent variable $g$. This independent variable $g$ represents the cumulative opacity distribution function and is then given by the opacity distribution function $f$, such that
\begin{equation}\label{eq:g}
	g(\kappa) = \int_0^\kappa f(\kappa^\prime) \mathrm{d}\kappa^\prime.
\end{equation}
The cumulative opacity distribution function $g$ can be understood as the probability to find an opacity value of less than $\kappa$ at a specific frequency $\nu$. While both Eqs.~\ref{eq:transm} and \ref{eq:transm_g} are formally identical, since they only differ in the order in which the sum is evaluated, they might differ significantly in the discrete limit, where individual summation points need to represent the value of the opacity for a certain nonzero width $\Delta\nu$ or $\Delta g$. In practice, the correlated-k method divides the total computed frequency range into coarse frequency bins, in which all integrals over frequency are substituted into integrals over $g$. The radiative transfer equation for the intensity $I(\nu,g)$ [\SI{}{\W\per\square\m\per\steradian\per\Hz}] of the coarse frequency bin between $\nu_0$ and $\nu_1$ at the discrete sub bin $g$ in $g$-space then becomes
\begin{equation}
	\mathbf{n}\cdot\nabla I(\nu,g) = \rho\kappa(\nu,g)\left[S(\nu,g)-I(\nu,g)\right],
\end{equation}
where $S(\nu,g)$ [\SI{}{\W\per\square\m\per\steradian\per\Hz}] is the source function and $\mathbf{n}$ is the unit vector of the direction in which the intensity is measured. Thus, in each of these coarse frequency bins, the radiative transfer is solved individually for each $g$ grid value and integrated over $g$ similarly to Eq.~\ref{eq:transm_g} afterward to obtain the intensity $I$ of the coarse frequency bin from $\nu_0$ to $\nu_1$. In this way, the correlated-k method allows for rapid calculations by requiring less radiative transfer computations for the same level of accuracy.

To obtain the correct k-tables for a mixture of individual species, one would need to sum up the individual contributions of the individual species, weighted with their abundance, and construct the opacity distribution functions of the total opacity. Several methods have been put forward to solve this issue, and we explain how we tested some of these in this work. It is important to note that the opacity distribution functions vary as a function of pressure and temperature and mixture in the gas, thus, in order to use the correlated-k method, one requires methods to construct these $\kappa(\nu,g)$ (k-table) values accurately. 

\subsection{Random overlap with rebinning and resorting (RORR)}\label{sec:RORR}
Both \citet{Saxner1984ODF} and \citet{Lacis1991ck} independently introduced a similar method for ODFs and k-tables respectively that treats the mixing of multiple opacity species under the assumption that the distributions of the individual opacity species are not correlated:
\begin{equation}
	f_\mathrm{tot}(\kappa_1, ... ,\kappa_{N_s}) = f(\kappa_1) \cdot ...\cdot f(\kappa_{N_s}).
\end{equation}
In simple terms, this means that, for example, the line cores are randomly distributed and do not systematically occur at the same frequencies. In the case of correlated-k, this method is called the random overlap with rebinning and resorting (RORR) method. An in depth introduction to the RORR method can be found in \citet{Amundsen2017ck} and in \citet{Molliere20151Dmodel}, and we instead just briefly outline its basic function. 

The core of the RORR method is the assumption that the opacities of two species are uncorrelated. This will then imply that their transmissions are also uncorrelated \citep{Molliere20151Dmodel}. From Eq. \ref{eq:transm}, we can then see that the transmission of both species can be multiplied to get the total transmission. It is then possible to find a k-table of a mixture by convolving their probability distributions \citep[e.g.,][]{Molliere20151Dmodel,Amundsen2017ck}. The final result of the convolution calculation can be resorted and binned back to the original $g$ grid for further computations. Repeating this procedure with the combined k-table of the two species and a third species yields the next step. This procedure is then repeated until all opacity species are included in the total k-table (\citet{Amundsen2017ck} provides a useful visualization of the procedure in their Fig.~1). From the methods outlined in this work, RORR is the slowest but most accurate method. Furthermore, RORR is well benchmarked against line-by-line calculations \citep[e.g.,][]{Amundsen2014GCM,Molliere20151Dmodel}. It is therefore the method of choice for most correlated-k 1D atmosphere models, such as \texttt{ATMO} \citep{Tremblin2015ATMO}, \texttt{petitRADTRANS} \citep{Molliere20191Dmodel} or \texttt{PICASO} \citep{Picaso1DRT}. 

\subsection{Premixed k-tables}\label{sec:pre}
Assuming equilibrium chemistry, one can create k-tables as lookup tables of pressure and temperature, which can be interpolated on during the radiative transfer calculations. It is important to note that these premixed tables are subject to the exact input to the equilibrium chemistry calculations (e.g., metallicity or C/O ratio) and need to be recomputed if the atmosphere is expected to deviate from these. Premixed tables can be computed in multiple ways. \citet{Showman2009GCM} calculated premixed tables by calculating the distribution functions of the mixture. In \citet{Schneider2022GCM}, we have calculated these k-tables using RORR on the individual k-tables. As already pointed out in \citet{Amundsen2017ck}, the accuracy of this approach is subject to the pressure and temperature resolution of the lookup table, since the equilibrium abundances are expected to vary by many orders of magnitude, much more than the individual opacities themselves. In \citet{Schneider2022GCM}, we have therefore computed the lookup tables, such that they match the pressure grid used in the GCM, removing the need to interpolate in pressure and allowing for a fine grid in temperature (1000 temperature points). While these lookup tables can be very precise, if resolved sufficiently, they come at the cost of flexibility, since they require assumptions on the abundances as a function of temperature and pressure.

\subsection{Summation}\label{sec:add}
In fact, the easiest way to calculate the mixed k-table from a mixture of different opacity species is to approximate the convolution by a sum:
\begin{equation}
	\kappa_\mathrm{tot}(\nu, g) = \sum_{i=1}^{N_s}\kappa_i(\nu, g),
\end{equation}
where $N_s$ is the number of species and the subscript $\kappa_i$ is the individual k-table of species $i$, weighted with its mass mixing ratio. While there is no logical justification for this approach, the approach is certainly the fastest method, as it only requires the evaluation of the sum of k-tables. This approach is certainly attractive, it will, however, naturally underestimate $\kappa_\mathrm{tot}$ at small $g$ and overestimate $\kappa_\mathrm{tot}$ at large $g$. This can be best seen in RORR\footnote{for a visualization, see Fig.~1 of \citet{Amundsen2017ck}}, where the evaluation of the convolution would add $\kappa_i$ values from larger $g$ as well as those of smaller $g$ to the $\kappa_\mathrm{tot}$ values at small $g$ (see Appendix~\ref{app:gval} for a quantitative comparison). This is particularly important, since the small $g$ values, which encode the small $\kappa_\mathrm{tot}$ values, decide the depth up to which stellar irradiation can be absorbed.

\subsection{Adaptive equivalent extinction}\label{sec:AEE}
Adaptive equivalent extinction (AEE) is a variation of equivalent extinction \citet{EdwardsSlingo1996RT} introduced by \citet{Amundsen2017ck}. The idea is to determine the most important species and then treat all other species as gray within the spectral band. To obtain the major absorber, one first calculates semi-gray opacities for each species. The semi-gray opacity of species $i$ is calculated as an average of the k-table in a given spectral bin and is weighted with a function $w$ that depends on the value of $g$:
\begin{equation}\label{eq:kappa_av}
	\kappa_{\mathrm{av},i}(\nu) = \frac{\int_0^1 \kappa_i (\nu,g)w(\nu,g)\mathrm{d}g}{\int_0^1 w(\nu,g)\mathrm{d}g}.
\end{equation}
We show below that the choice of the weighting function $w$ has a significant impact on the accuracy of the AEE method. A good measure for the importance of the opacity at a given $g$ value on the accuracy of the radiative transfer calculation is the magnitude of the flux at that value of $g$ \citep{Amundsen2017ck}. We thus chose to use the absolute values of the stellar and planetary fluxes through a $g$ value as a weighting function. However, since we can only know the fluxes, once we have already mixed the k-tables and calculated the radiative transfer, we need to rely on the value of the planetary and stellar flux from a previous radiative transfer calculation, which would be the previous radiative time step in the case of GCMs.

Using these $\kappa_{\mathrm{av},i}(\nu)$ values, a major absorber is found by vertically integrating the transmission (Eq. \ref{eq:transm}) from the top of the atmosphere down to an optical depth of one for each species. The first species to reach an optical depth of one is then used as the major absorber in the vertical column. The final total opacity in each spectral bin is then given as 
\begin{equation}
	\kappa_{\mathrm{tot}}(\nu, g) = \kappa_m(\nu,g) + \sum_{i\ne m}^{N_s} \kappa_{\mathrm{av},i}(\nu),
\end{equation}
where $m$ is the species, which has been identified as the major absorber. We note that the UM uses a slightly different and less sophisticated method (called equivalent extinction or EE), where the major absorber is determined locally and without integrating over the atmospheric column. A detailed introduction of the AEE method can be found in \citet{Amundsen2017ck}.

\subsection{DeepSet approach}\label{sec:deepset}
The RORR mixing approach has three important attributes. Firstly, the method stays the same, no matter how many species are mixed with each other and in what order. RORR is thus invariant to permutations in the set of opacities that are to be mixed. Secondly, to a first approximation RORR can be approximated by a simple sum as mentioned in Sect.~\ref{sec:add} and verified below. Lastly, although RORR is computationally expensive compared to the other outlined approaches, in its core for each frequency bin, it only consists of a convolution, a sorting step and an interpolation step, which is repeated $N_s-1$-times. These three attributes greatly constrain a possible emulation of RORR by a machine learning algorithm, since only algorithms with fast inference, versatile input size and permutation invariance can be used.

We tested several architectures, such as a convolutional network similar to the U-net \citep{Ronneberger2015Unet}, gradient boosted regression trees using the abundances as input and the mixed opacities as output using \texttt{XGboost} \citep{Chen2016XGBoost}. With the U-net we ended up needing too many convolution blocks and with \texttt{XGboost}, we needed structures that were too deep and therefore too memory consuming to get reasonable accuracy. One of the reasons for the poor performance of these methods in our context is that they are too different from a simple sum (see Sect.~\ref{sec:add}). We have therefore settled with a \texttt{DeepSet} approach \citep{Zaheer2017DeepSets}. A \texttt{DeepSet} for our case of k-table mixing can be written as 
\begin{equation}\label{eq:ds}
	y = \aleph\left(\bigoplus_{i=1}^{N_s} \beth\left(X_i\right)\right),
\end{equation}
where $y$ is the response of the \texttt{DeepSet} (the mixed k-tables, see below Eq.~\ref{eq:y}), $X=\{X_1, ..., X_{N_s}\}$ is the set of input vectors (the individual k-tables, see below Eq.~\ref{eq:x}), and $\aleph$ and $\beth$ are functions.  

\begin{figure}
	\centering
	\includegraphics[width=.45\textwidth]{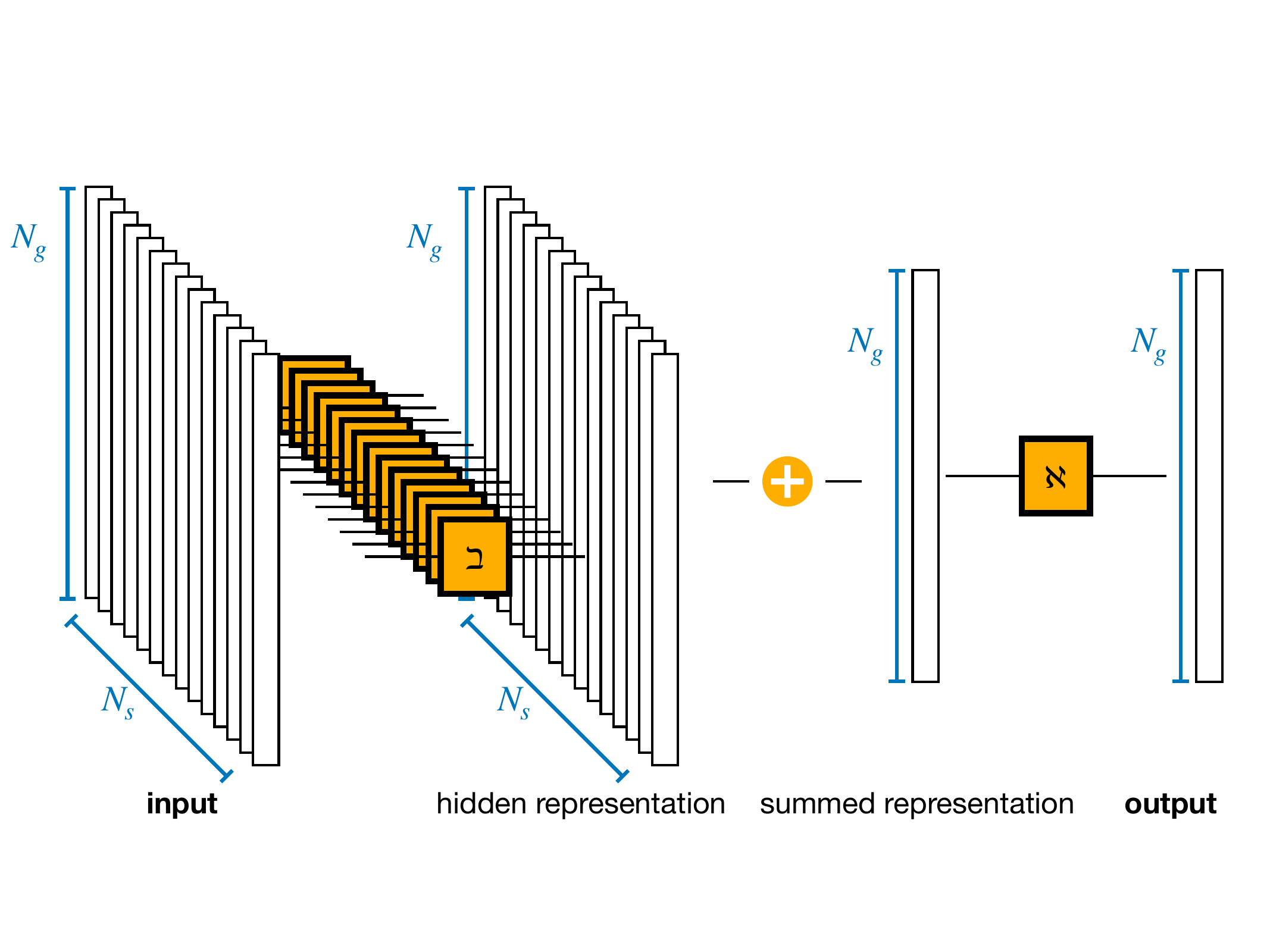}
	\caption{Neural network used in the DeepSet (DS) approach.}
	\label{fig:deepnet}
\end{figure}

Simply put, the idea is to use a function $\beth$ to encode each vector $X_i$ of a set into a hidden representation $\beth(X_i)$. These representations are then summed up\footnote{Note that any permutation invariant operation could be used in this step.}, and subsequently decoded by function $\aleph$ to get the output. We illustrate the concept of \texttt{DeepSets} \citep{Zaheer2017DeepSets} for the problem of k-table mixing in Fig.~\ref{fig:deepnet}. We used
\begin{equation}
	\beth(X_i) = \max(a_1 \cdot X_i, 0)
\end{equation}
and 
\begin{equation}
	\aleph(z) = a_2 \cdot z,
\end{equation}
with weights $a_1$ and $a_2$, which are matrices of size $N_g \times N_g$. Whereas $\aleph$ is a linear function, $\beth$ is nonlinear due to the inclusion of a rectified linear activation (ReLU). Different, more complex functions could also be used, but we found that the accuracy reached by using these simple functions is already good enough. More complex functions, would thus only result in a slower performance. The weights $a_1$ and $a_2$ of the functions $\aleph$ and $\beth$ are then learned by a neural network.

To train the neural network, we chose to implement the network in \texttt{Keras} \citep{chollet2015keras}. The weights are learned by minimizing the mean squared error using the Adam optimizer \citep{Kingma2014ADAM}. We then performed a Bayesian hyperparameter search using \texttt{hyperopt} \citep{Bergstra2012Hyperopt} to find the best amount of features for the hidden representation and to determine the optimal learning rate of the optimizer. The loss did not improve significantly by the use of more than $N_g$ features in the hidden representation, and we therefore chose to use $N_g$ features. Additionally, we found that a learning rate of $\alpha=\num{1e-3}$ seems to perform best.

The neural network acts on each frequency bin individually and therefore does not care about the frequency resolution. To generate the training and test data, we computed the mixed k-tables with RORR from $\approx \num{8e5}$ sets of 14 k-tables each. These 14 k-tables were taken from the 11-bin resolution (S1) of the 14 opacity species taken into account in \texttt{expeRT/MITgcm} \citet{Schneider2022GCM} (see Sec.~\ref{sec:methods} for details on the species) and were uniformly randomly weighted with reasonable abundance ranges.\footnote{We note that the network can, by construction, deal with any number of species without retraining needed}. The input for the network is not the plain individual k-tables, but instead we scale them with the sum of the k-tables as
\begin{equation}\label{eq:x}
	X_i(\nu, g) = \log \left(\frac{\kappa_i(\nu,g)}{\sum_{j=1}^{N_s}\kappa_j(\nu, g)}\right).
\end{equation}
Similarly, we scale the targets (e.g., the predictions of the network) by
\begin{equation}\label{eq:y}
	y(\nu, g) = \log \left(\frac{\kappa_\mathrm{tot}(\nu,g)}{\sum_{j=1}^{N_s}\kappa_j(\nu, g)}\right).
\end{equation}
The advantage of this input and output scaling comes three-fold. Foremost, we are interested in minimizing the error of the small values in the k-table, as those are the ones that generate windows in the spectrum, which are very important for the temperature structure, hence taking the logarithm is useful because it pronounces relatively large deviations of small k-table values in the loss. Secondly, we find that the sum of the k-tables is already a reasonable approximation for the mixed k-tables (see Sec.~\ref{sec:results}). Thirdly, the output scaling nicely captures the positivity of the problem, prohibiting the possibility of negative predictions when reversing the output scaling. It is also important to stress that all species $\kappa_i(g)$ are fed individually through the same function $\beth$ (with the same weight) to create a unique nonlinear representation for each species.

The main advantage of the DeepSet method is its flexibility, because trained on the individual frequency bins of k-tables, it operates independent of chemical composition or opacity species. It can by construction of the training set operate on any composition and metallicity. Furthermore, we think that it can in principle work on any set of k-tables with shapes that are similar enough to those of the training set. We have tested this by changing the frequency resolution of the training set. Doing so, we found that this did not significantly affect the accuracy, when applied to a different frequency resolution than the one being trained on. We therefore think that it would be only necessary to retrain the network, if the discretization of $g$ values changes. We discuss further numerical considerations of this mixing method in Appendix~\ref{app:num}.

\section{Methods}\label{sec:methods}
\begin{table*}
	\centering
	\caption{Simulations}
	\begin{tabular}{lcc}
		\hline\hline
		label & mixing method & reference\\
		\hline
		RORR & random overlap with rebinning and resorting (RORR)& Sect.~\ref{sec:RORR}\\
		PRE & premixed k-tables & Sect.~\ref{sec:pre}\\  
		DS & DeepSet approach & Sect.~\ref{sec:deepset}\\
		AEE\_we & adaptive equivalent extinction with flux weighting & Sect.~\ref{sec:AEE}\\
		AEE & adaptive equivalent extinction without flux weighting & Sect.~\ref{sec:AEE}\\		
		ADD & sum of all k-tables & Sect.~\ref{sec:add}\\	
		\hline			
	\end{tabular}
	\begin{tablenotes}
		\item \textbf{Notes:} Explanation of the mixing methods used in the individual simulations, as labeled in the figures of this work.
	\end{tablenotes}
	\label{tab:labels}
\end{table*}

To test the individual mixing methods, we use the 3D GCM \texttt{expeRT/MITgcm} \citep{Carone2020GCM, Schneider2022GCM}. \texttt{expeRT/MITgcm} builds on the dynamical core of the \texttt{MITgcm} \citep{Adcroft1997MITgcm,Adcroft2004}, which solves the hydrostatic primitive equations of hydrodynamics on a cubed-sphere grid. In order to accurately account for radiative heating and cooling, \texttt{expeRT/MITgcm} solves the radiative transfer using the Feautrier method \citep{Feautrier1964} with approximate Lambda iteration \citep{Olson1986} and Ng-acceleration \citep{Ng1974}. The routine that solves the radiative transfer is an adapted version of the flux routine in \texttt{petitRADTRANS} \citep{Molliere20191Dmodel,Molliere20201Dmodel}. We have incorporated the radiative transfer solver and benchmarked it in \texttt{expeRT/MITgcm}. We found in \citet{Schneider2022GCM}, that the combination of five frequency bins and 16 $g$ values achieves good enough accuracy for long term convergence studies such as those in \citet{Schneider2022GCMLetter} and \citet{Sainsbury-Martinez2023Inflation}. In this work, we use 11 frequency bins, with each 16 $g$ values, which is good enough for comparisons of GCMs to observations. We note that other hot Jupiter GCMs typically use eight $g$ values, with sometimes a higher frequency resolution of $\approx 30$ frequency bins \citep{Showman2009GCM,Amundsen2016UKMetGCM,Lee2021GCM}. Future studies should investigate whether using fewer frequency points and more $g$ values is more accurate than using fewer $g$ values and more frequency points. The line opacity species used in this work are identical to the ones used in \citet{Schneider2022GCM} and include H$_2$O \citep{Polyansky2018Opac}, CO$_2$\citep{Yurchenko2020Opac}, CH$_4$ \citep{Yurchenko2017Opac}, NH$_3$ \citep{Coles2019Opac}, CO \citep{Li2015Opac}, H$_2$S \citep{Azzam2016Opac}, HCN \citep{Barber2014Opac}, PH$_3$ \citep{Sousa-Silva2015Opac}, TiO \citep{McKemmish2019Opac}, VO \citep{McKemmish2016Opac}, FeH \citep{Wende2010Opac}, Na \citep{Piskunov1995Opac}, and K \citep{Piskunov1995Opac}. 

We chose to use HD~209458 b as a planet and the setup is identical to the setup in \citet{Schneider2022GCM}, where the models in this work only differ by the method with which opacities are mixed. The different mixing methods and their corresponding labels are laid out in Table~\ref{tab:labels}. To this end, we have implemented each of the abovementioned mixing methods. \texttt{expeRT/MITgcm} can now run in two modes by either mixing k-tables during runtime (utilizing one of the aforementioned methods) or using premixed k-tables. To incorporate mixing during runtime in the GCM, we updated our preprocessing toolkit to additionally output a pressure-temperature grid of equilibrium abundances (taken from the \texttt{easyCHEM} \citep{Molliere20171Dmodel} interface to \texttt{petitRADTRANS} \citep{Molliere20191Dmodel}), along with a pressure-temperature grid of k-tables for the individual absorbers. In the on-fly mixing mode, abundances and k-tables of each of the considered absorbers are linearly interpolated to the pressure and temperature in the GCM, weighted by their abundance and then mixed by one of the abovementioned methods. 

The weighting in the adaptive equivalent extinction method induces the need for more scattering iterations, because the k-table becomes dependent on the bolometric flux from the previous time-step, effectively inserting a time dependent perturbation into the opacities, because the scattering source function will be subject to these opacity perturbations as well, rendering its guess from the previous time-step less accurate. For performance reasons, we thus found that the weighted adaptive equivalent extinction method required us to limit the amount of maximum scattering iterations per radiative time step to two instead of 500, which is generally enough for planets similar to HD~209458 b with only Rayleigh scattering, since the source function is reused as initial guess in the next radiative time-step \citep[see][for an explanation of scattering in \texttt{expeRT/MITgcm}]{Schneider2022GCM} and the source function is thereby naturally iterated on during model convergence. To be consistent in all models, we have thus chosen to generally limit the amount of scattering iterations per radiative time-step to two, if not otherwise stated.

All models have been integrated up to 2000 days with a radiative time-step of \SI{100}{\s} and a dynamical time-step of \SI{25}{\s}, which are typical values for hot Jupiter GCMs \citep[e.g.,][]{Showman2009GCM,Lee2021GCM,Schneider2022GCM}. All models use equilibrium chemistry to constrain the abundances. In practice, the code interpolates the abundances on a grid of pressure and temperature. In Section \ref{sec:rain}, we show a model, where we use the DeepSet mixing of k-tables, but removed all of TiO and VO, whenever TiO and VO would reappear in equilibrium chemistry in the gas phase, although it is condensed out further down in the atmosphere. This method is similar to the methods of rainout described elsewhere \citep[e.g.,][]{Lodders2002rainout,Marley2021BrownDwarf}, but less sophisticated compared to 3D models that include proper chemical transport schemes \citep[e.g.,][]{Parmentier2013coldtrap,Lee2023GCMChem,Drummond2018GCMChem}. A more detailed description of the algorithm for the detection of rainout is outlined in Appendix~\ref{app:rainout}.

\section{Results}\label{sec:results}
In this work, we compare the mixing methods introduced in Sect.~\ref{sec:mixing_intro} to the slow RORR method. An additional comparison between the premixed method (PRE) used in our GCM, as introduced and used in \citet{Schneider2022GCM}, and the RORR method can be found in Appendix~\ref{app:prevsrorr}. Mixing opacities during runtime in a GCM requires a tradeoff between accuracy, performance, and flexibility. When comparing the accuracy of mixing methods, it is important to keep in mind that the low spectral resolution of the k-tables used in GCMs induces an error of a few percent on bolometric fluxes \citep[e.g.,][]{Amundsen2014GCM,Leconte2021exok,Schneider2022GCM}. It is therefore pointless to aim for accuracies of less than one percent, since the overall error will be governed by the chosen spectral resolution. Comparing the individual mixing methods with each other thus needs to consider all of these perspectives. In order to have a fair comparison between the individual methods, we chose to compare all simulations in two aspects: The accuracy on the resulting atmospheric state and the computational time. We provide additional accuracy diagnostics in Appendix~\ref{app:gval}, where we compare the $\kappa_\mathrm{tot}$ values obtained by different methods. We also diagnose fluxes and heating rates in Appendix~\ref{app:acc}.

\begin{figure*}
	\centering
	\includegraphics{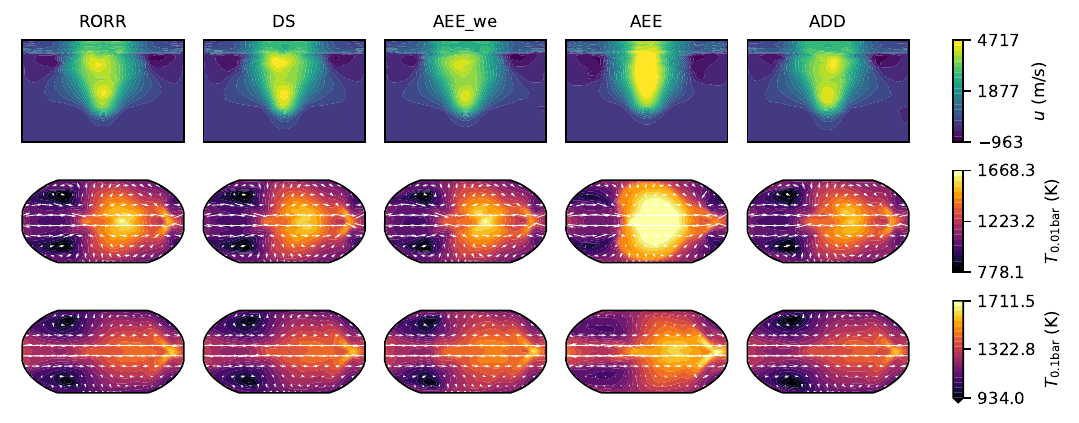}
	\caption{Zonally averaged eastward winds (first row) and temperature slices at \SI{0.01}{\bar} (second row) and \SI{0.1}{\bar} (third row) for the different models considered. All colors are normalized to the first column (RORR).}
	\label{fig:winds+temps}
\end{figure*}
\begin{figure*}
	\centering
	\includegraphics{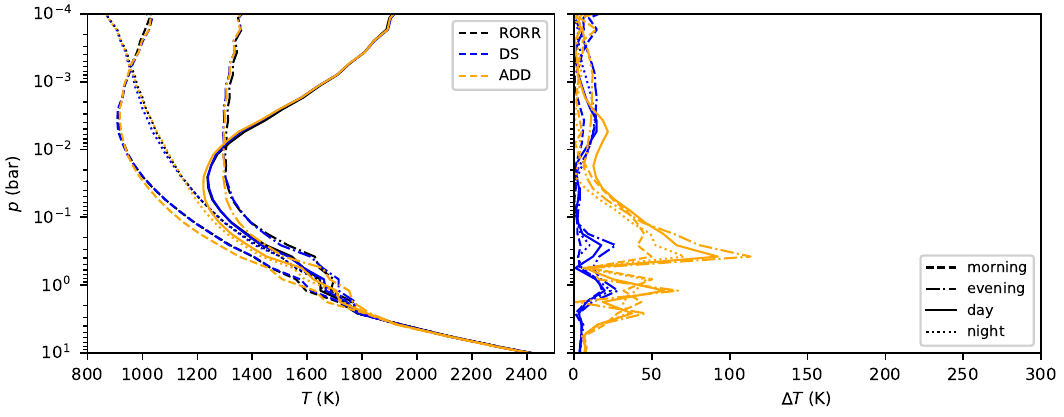}
	\caption{Temperature profiles for different parts of the atmosphere of the RORR, DS, and ADD simulations from Fig.~\ref{fig:winds+temps}. The different colors represent different simulations, whereas the different line-types represent different parts of the atmosphere. The left panel shows the temperature profile and the right panel shows the absolute difference between the temperature profile to the temperature profile of the RORR simulation.}
	\label{fig:RORR_ML_ADD}	
\end{figure*}
\begin{figure*}
	\centering
	\includegraphics{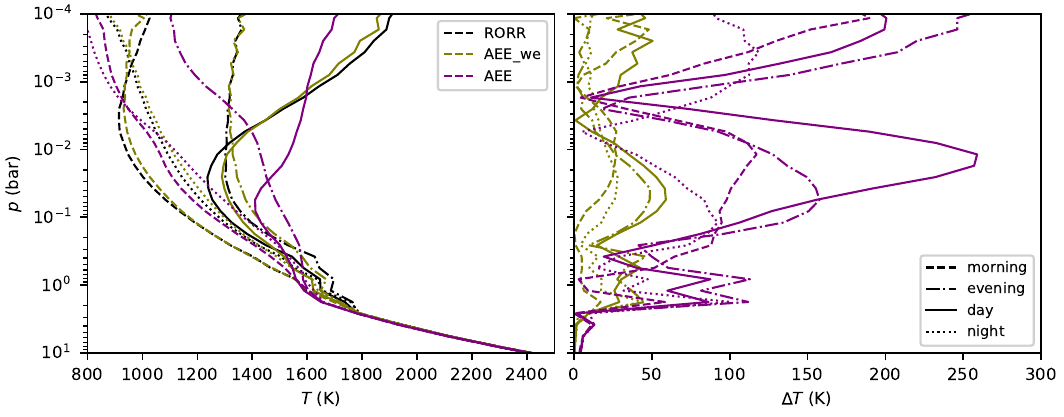}
	\caption{Temperature profiles for different parts of the atmosphere of the RORR, AEE, and weighted AEE (AEE\_we) simulations from Fig.~\ref{fig:winds+temps}. The panels and their meaning are identical to Fig.~\ref{fig:RORR_ML_ADD}.}
	\label{fig:RORR_AEE}
\end{figure*}

In order to qualitatively compare the mixing methods, we show the atmospheric state at two pressure layers (\SI{0.01}{\bar} and \SI{0.1}{\bar}) together with the zonal wind speed in Fig.~\ref{fig:winds+temps}. Aside from the adaptive equivalent extinction (AEE) model without flux weighting, all models look similar at a first glimpse. The jet strength and day-night temperature contrasts are not significantly affected by these different methods. However, the AEE method seems to produce a significantly higher day-night temperature contrast and a faster jet. These differences are more pronounced at lower pressure, but are still notable at higher pressures of \SI{0.1}{\bar}. 

Since most methods produce the same qualitative trend in winds and temperatures, it can be useful to calculate spatial temperature averages and compare those to the RORR models. For visual reasons, we have split these comparisons into two figures, where Fig.~\ref{fig:RORR_ML_ADD} compares the DeepSet (DS) and summation (ADD) method to the RORR method and Fig.~\ref{fig:RORR_AEE} compares both the weighted and non-weighted adaptive equivalent extinction methods to the RORR method. It was surprising to see that the ADD method performs well, given its simplicity and methodological flaws. However, the temperature is often slightly cooler at higher pressures, which might be related to the underestimation of $\kappa_\mathrm{tot}$ at small $g$ and the overestimation at large $g$ (see Appendix~\ref{app:gval}). The overestimation of $\kappa_\mathrm{tot}$ at large $g$ can lead to an enhancement of the absorption of the stellar flux in the upper layers, which cannot penetrate deep enough to cause heating in the deeper layers. These flaws of the ADD method do not seem to persist in the DS mixing, which uses the ADD method in its preprocessing (see Sect.~\ref{sec:deepset}). Instead, we find that the DS mixing performs very well.

Looking at the adaptive equivalent mixing method with and without weighting, we find that the AEE\_we method is almost as accurate as the DS mixing. The weighting certainly helps to find a good estimate of the major absorber and drastically increases the accuracy of this approach. Looking at the residuals, we see a strong correlation between the error of the weighted and non-weighted method, which points to a general issue of the method instead of an issue with the major absorber. Unlike in the case of simply summing up k-tables, the AEE method tends to not overestimate $\kappa_\mathrm{tot}$ at large $g$ but instead to underestimate it. This might be explained by the minor absorbers, which flatten a k-table by offsetting $\kappa_\mathrm{tot}$ at the small $g$ and decreasing the impact of the high $g$ values (see Appendix~\ref{app:gval}). Similar to the overestimation of $\kappa_\mathrm{tot}$ at small $g$ values, the underestimation of $\kappa_\mathrm{tot}$ at large $g$ values also shifts the location at which irradiation is absorbed, in this case, into the opposite direction by leading to less absorption in the uppermost layers. 

We thus conclude that both the ADD method and AEE method, with and without weighting, introduce systematic noise to $\kappa_\mathrm{tot}$ at both small and large $g$ values. This noise can amplify the errors of the AEE and ADD methods. In contrast, the DS method exhibits no systematic error (see Appendix~\ref{app:gval}) but instead uniformly distributed random noise. The overall error in temperature estimation is thereby not significantly affected. Therefore, we do not recommend using the ADD method or unweighted AEE method and instead recommend the use of the DS method.

\begin{table}
	\centering
	\caption{Runtime}
	\begin{tabular}{lcc}
		\hline\hline
			label & time [\SI{}{\hour}] & relative to PRE\\
			\hline
			PRE & 3.91 & 1.00 \\
			ADD & 7.20 & 1.84\\
			AEE & 7.44 & 1.90\\
			AEE\_we & 13.37 & 3.42\\									
			DS & 10.88 & 2.78\\
			RORR & 25.12 & 6.43\\			
			\hline
	\end{tabular}
	\begin{tablenotes}
		\item \textbf{Notes:} Runtime of the GCM needed for the first 100 days of the simulation. The node used to run the model includes 2x Intel(R) Xeon(R) Gold 6248R CPU @ 3.00GHz, and we utilized all 48 cores.
	\end{tablenotes}
	\label{tab:ctime}
\end{table}

In terms of computational costs, one needs to consider two general computational overheads during runtime, compared to using a premixed grid. The first overhead comes from the handling of the individual k-tables, such as the interpolation, as compared to handling of just one premixed k-table. Secondly, obviously the computation of the mixing itself. We show the computation time needed to run the initial 100 days of the simulation in Table~\ref{tab:ctime}. The performance of the summation (ADD) method, in which individual k-tables are simply added up, is mainly constrained by the handling of the individual k-tables, since the cost of the summation can be neglected, whereas all the other methods are also subject to the computational cost of the mixing. Therefore, a significant fraction of the computational cost in the AEE, AEE\_we, and DS simulations can be explained by the overhead of handling individual k-tables. 

The high computational cost of RORR mixing makes it impossible to use it during model runtime. Even when using an optimized sorting algorithm, which is now the standard in \texttt{petitRADTRANS}, and which significantly speeds up RORR, the RORR technique performs at least six times slower compared to the premixed case. One of the main reasons for the poor performance of RORR in our specific setup is the quadratic computational dependence on the number of $g$ values. GCMs that use eight $g$ values instead of 16 could therefore (at the cost of accuracy) have faster performance of the RORR method \citep[for a discussion see][]{Amundsen2017ck}.

Due to the increased amount of scattering iterations, the adaptive equivalent extinction with weighting (AEE\_we) is generally the slowest of all the approaches. The poor convergence behavior of the AEE\_we method makes this method less reliable and less performant. This will be an even bigger issue, when scattering becomes non-negligible, which could be the case for lower temperatures or if clouds and hazes are included. However, when limiting the maximum amount of scattering iterations to two, as we have done here, it is fast enough to compete with the other methods and if the weighting were to be neglected (AEE), it would even be similarly fast as the ADD method and faster than the DS mixing. Performance-wise, we thus think that either of the ADD, AEE (with and without flux weighing), or DS mixing approach could be equally used in a setup that requires mixing during runtime.

Based on the discussions and findings above and in Appendices~\ref{app:gval} and \ref{app:acc}, we conclude that the AEE\_we or DS method should be used in a GCM for accurate results with good computational performance. However, in our multi-stream setup where we iterate over the source function, we do not recommend using the AEE\_we method due to numerical issues with its convergence.

\section{Rainout}\label{sec:rain}
\begin{figure}
	\centering
	\includegraphics{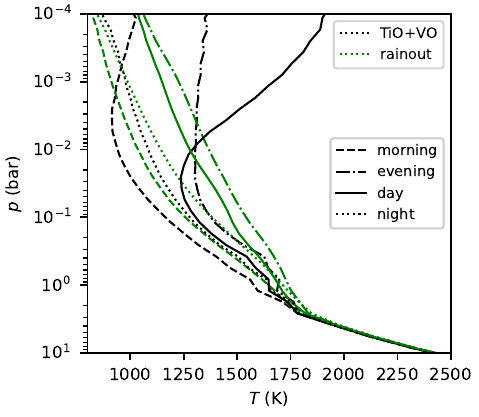}
	\caption{RORR model from Fig.~\ref{fig:winds+temps} compared to a model with rainout of TiO and VO. The stratosphere disappears if rainout is included in the model.}
	\label{fig:rain}
\end{figure}
Mixing during runtime is only relevant in GCMs if the chemical abundances are to be changed from chemical equilibrium. A sufficiently well resolved premixed table, will in most cases of chemical equilibrium be the fastest and most reliable method of choice. However, in the case of chemical disequilibrium, which is the case if photochemistry or chemical kinetics were to be considered, we can not simply premix k-tables, because the abundances of the relevant opacity species need to be changed during runtime. Another very simple scenario for such a situation could be the rainout of heavy refractory species such as TiO and VO. Due to their strong absorption in the UV, they absorb a significant fraction of the stellar flux at high altitude, thus heating the upper atmosphere significantly. Such a strong heating in the upper atmosphere can lead to a thermal inversion, where the atmosphere becomes hotter toward the top  \citep[e.g.,][]{Showman2009GCM}. However, due to vertical mixing and advection, condensed TiO and VO could gravitationally settle and therefore not be available in the gas phase at higher altitudes \citep[e.g.,][]{Parmentier2013coldtrap}.

In Fig.~\ref{fig:rain} we show a model, in which all TiO and VO is removed by a simple rainout prescription (see Appendix~\ref{app:rainout} for details), which removes all TiO and VO from the gas phase, if it is condensed further down in the atmosphere. Using a premixed table, we would not be able to calculate the radiative effect of the rainout on the atmospheric structure, however, by using mixing during runtime we can trace the effect of the change in chemical abundances on the temperature. The lack of the strong UV absorbers TiO and VO in the upper layers means that the upper layers get cooler, because less of the stellar flux is absorbed in those layers, whereas the intermediate pressure layers get warmer, where the bulk of the stellar flux is absorbed instead. As expected, we find that the thermal inversion caused by TiO and VO, as seen for example in Fig.~\ref{fig:RORR_ML_ADD}, is self-consistently removed. The absence of the stratosphere on the day side effects the night side as well. This effect is noticeable at \SI{0.1}{\bar}. In the simulation with rainout, the absence of the stratosphere leads to warmer gas compared to the simulation without rainout. As a result, the superrotating jet transports the warmer gas to the nightside, causing it to heat up.


\section{Discussion and conclusion}\label{sec:discussion}
The correlated-k method is a useful approximation for rapid radiative transfer calculations with accuracies of a few percent \citep[e.g.,][]{Amundsen2014GCM,Leconte2021exok,Schneider2022GCM}, when used with resolutions typical for GCMs. We have demonstrated the performance and accuracy of several methods that could be used in GCMs to calculate the total opacity in the correlated-k assumption. We extended the work of \citet{Amundsen2017ck}, who performed a similar analysis for the adaptive equivalent extinction (AEE) method. Furthermore, we have introduced two additional methods: The DeepSet (DS) mixing and a simple method in which k-tables are simply summed up (ADD). Whereas the work of \citet{Amundsen2017ck} only considered the accuracy of heating rates and fluxes, we incorporated the RORR, ADD, DS, and AEE mixing methods into our GCM to demonstrate the performance in a real application. The DeepSet method turns out to be fairly accurate and flexible, leveraging machine learning, to calculate k-tables of gas mixtures. Overall, we find that
\begin{enumerate}
	\item The random overlap with resorting and rebinning (RORR) method is too slow to be used in GCMs for mixing during model runtime.
	\item The AEE method requires a proper weighting to be accurate. Such a weighting, however, affects the numerical stability of the radiative transfer calculation, which will be especially important if scattering is non-negligible.
	\item The ADD method and the AEE method are prone to systematic errors. This is especially problematic for the unweighted AEE and the ADD method, rendering a use of these methods questionable.
	\item The DS method has minor statistical errors on fluxes and heating rates that do not seem to enhance the overall error, which seems to be an advantage of the DS method.
\end{enumerate}

The DS mixing method provided by this work is accurate and open source\footnote{\url{https://github.com/AaronDavidSchneider/opacmixer}}, and can be easily implemented in any radiative transfer package with no need to use complex libraries, as it only requires two matrix multiplications. Once trained, the network can perform on any composition and any set of opacity species. Although not strictly needed, we recommend training the network for a specific frequency resolution to maximize the accuracy. The amount of training data needed is small, and training can be performed within minutes on a standard personal computer. The provided open source package currently works with binned down \texttt{petitRADTRANS}-format k-tables, but can be easily extended to load any k-table format, and documentation is provided for how to achieve this. 

The methods tested in this work, have been tested in terms of accuracy on the atmospheric structure, and we think that these methods provided here will be key, when moving forward toward self-consistent transport of chemicals in the atmospheres of planets. We note, however, that these methods, do not translate to models that predict spectra. Future work is thus needed to test, if similar methods could also be used for atmospheric retrievals. 

By implementing a simple chemical rainout procedure, we mimic the gravitational settling of TiO and VO, to demonstrate the DS method in a use-case of disequilibrium chemistry. By accounting for rainout in this way, we find that TiO and VO can be trapped in the deeper atmosphere, thus hindering the formation of a stratosphere. We note, that this approach is fairly simplified, and hope that our work enables future models to treat cold trapping self-consistently.

\begin{acknowledgements}
A.D.S., L.D., U.G.J. and C.H. acknowledge funding from the European Union H2020-MSCA-ITN-2019 under Grant no. 860470 (CHAMELEON). U.G.J. acknowledges funding from the Novo Nordisk Foundation Interdisciplinary Synergy Program grant no. NNF19OC0057374. The bibliography of this publication has been typesetted using \texttt{bibmanager} \citep{bibmanager}\footnote{\url{https://bibmanager.readthedocs.io/en/latest/}}. The post-processing of GCM data has been performed with \texttt{gcm-toolkit} \citep{gcm_toolkit} \footnote{\url{https://gcm-toolkit.readthedocs.io/en/latest/}}. We would like to express our gratitude to the referee, Joanna Barstow, for her valuable comments that enhanced the quality of the manuscript.
\end{acknowledgements}

\bibliographystyle{aa}
\bibliography{ms}

\begin{appendix}
\section{Accuracy on individual $g$-values}\label{app:gval}
\begin{figure*}
	\centering
	\includegraphics{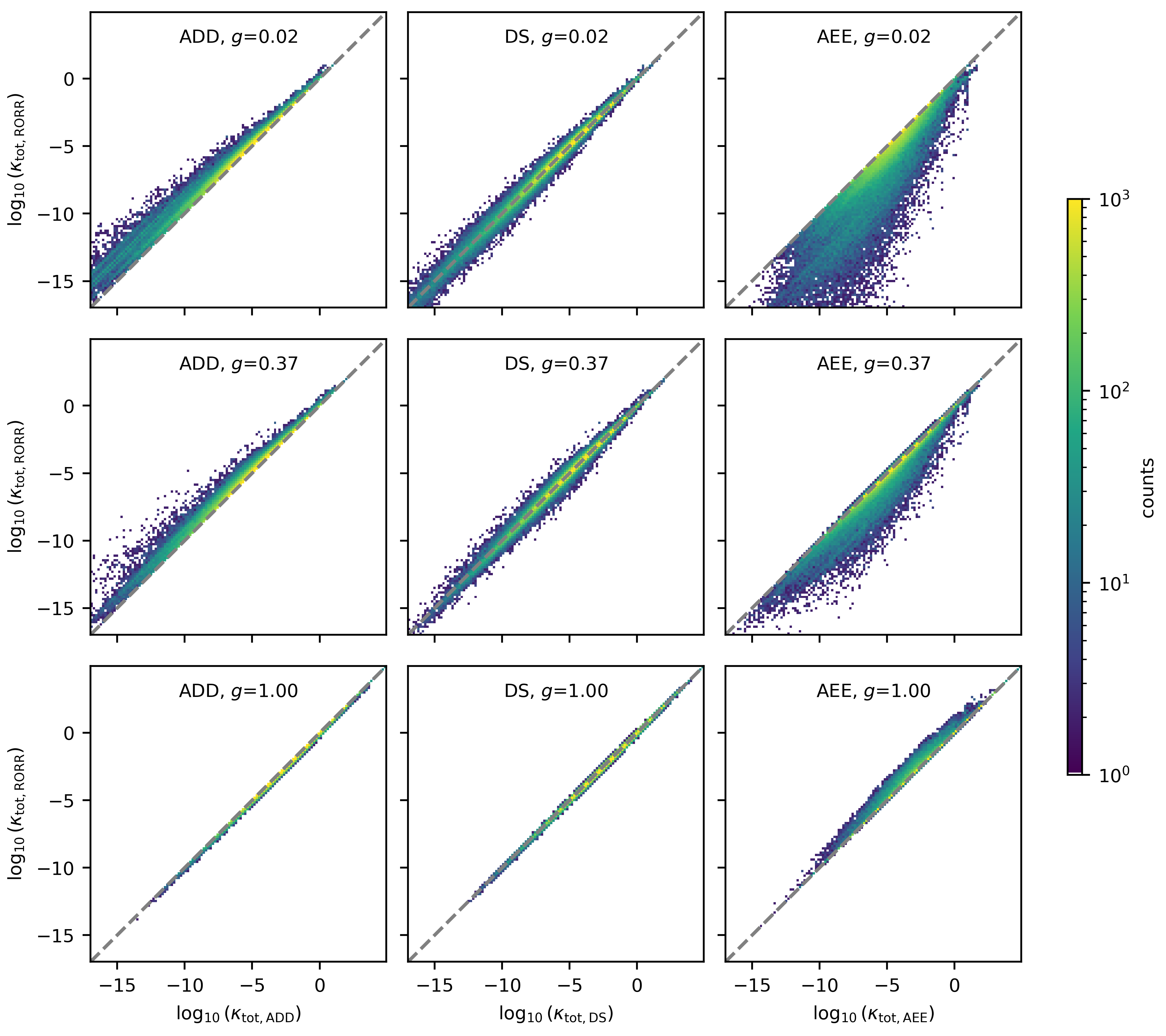}
	\caption{Mixed $\kappa_\mathrm{tot}$ values as a result of using different methods (x-axis, columns) compared to the respective values using the RORR method (y-axis) for three different $g$ values (rows). The data points are taken from the counterpart of the training set, used to train the DS method (see Sect.~\ref{sec:deepset}). The density of points is logarithmically color-coded.}
	\label{fig:g_predictions_density}
\end{figure*}
In Sect.~\ref{sec:AEE}, we mentioned that summing up averaged k-tables to the k-table of the major absorber in the AEE method leads to underestimation of high $g$ values and overestimation of small $g$ values. In contrast, as discussed in Sect.~\ref{sec:add}, the ADD method underestimates small $g$ values and may overestimate large $g$ values. On the other hand, the DS method fits the RORR method using mean squared loss, without favoring over- or underestimation of any $g$ value.

To test these methods for systematic errors, we generate mixed k-tables using the counterpart of the training set from Sect.~\ref{sec:deepset}. These k-tables are then compared to the results from the RORR method. The comparison is shown in Fig.~\ref{fig:g_predictions_density}. We note that the AEE method (both weighted and non-weighted) necessitates a 1D column to identify the maximum absorber (see Sect.~\ref{sec:AEE}). However, as our test is 0D, we were unable to implement this here. Instead, we determine the primary absorber in each set by selecting the highest value from the 14 k-tables.

In Fig.~\ref{fig:g_predictions_density}, we observe that the resulting $\kappa_\mathrm{tot}$ values from the DS method are symmetrically distributed around the values predicted from the RORR method, indicating a statistically random error. However, this is not the case for the AEE and ADD methods. More specifically, we find that the results for the AEE method indeed lead to a significant overestimation of $\kappa_\mathrm{tot}$ at small $g$ values and a clear underestimation of $\kappa_\mathrm{tot}$ at large $g$ values. However, we note again, that this trend would be less dramatic if the major absorber is picked wisely and if the averaged kappa values are determined from a proper weighting. It is thus vital to use the AEE method with a proper weighting, even though this might result in numerical challenges. Similarly, we see that indeed the ADD method leads to a significant underestimation of $\kappa_\mathrm{tot}$ at small $g$ values. The overestimation at large $g$ values, on the other hand, is very subtle and can only be seen by looking closely at the bottom left panel.

\FloatBarrier  
\section{Numerical considerations for the DeepSet implementation}\label{app:num}
One of the advantages of the DeepSet is its simplicity, which allows an easy naive implementation, since it only requires two matrix multiplication operations and one summation. There are several considerations regarding the performance of this approach. The first consideration is the use of the logarithm and, to reverse back to k-tables, the use of the exponential function. Both of these operations are unfortunately quite slow, albeit being needed as described in Sect.~\ref{sec:deepset}. Profiling the FORTRAN code used in the GCM shows that approximately half of the CPU time is spent on calculating the logarithm for the input scaling\footnote{The input scaling needs to be calculated $N_s$ times more often than the reverse output scaling, see Fig.~\ref{fig:deepnet}}. The other half of the CPU time is spent on the matrix multiplication, which can be optimized using hardware dependent compiler optimizations. For our architecture (using \texttt{ifort} on an intel CPU), we found that the build in \texttt{MATMUL} delivers the best performance compared to \texttt{MKL DGEMM}, \texttt{OpenBLAS DGEMM} and a manual implementation. The reason for the missing performance increase from the highly optimized \texttt{DGEMM} can most likely be found in the overhead of calling \texttt{DGEMM} compared to the small size of the matrix that is to be multiplied ($16\times16$).\footnote{We note that faster and more efficient implementations could also be achieved by stacking the computations and deploying the computations to a GPU.}

\FloatBarrier
\section{Benchmarking RORR against PRE}\label{app:prevsrorr}
\begin{figure*}
	\centering
	\includegraphics{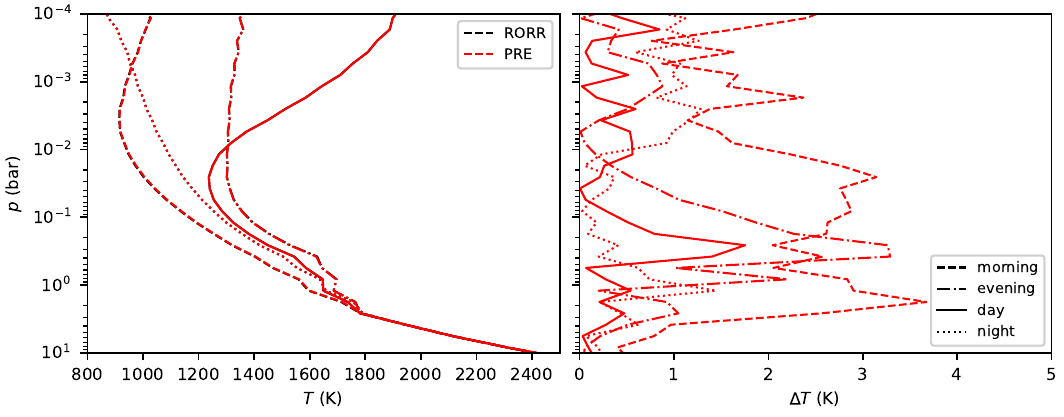}
	\caption{Temperature profiles of the RORR simulation compared to a simulation with premixing (PRE). The left panel shows the temperature profiles of both the RORR and PRE simulations and the right panel shows the difference between the RORR and premixed temperature profiles. The figure is in the same style as Figs.~\ref{fig:RORR_ML_ADD} and \ref{fig:RORR_AEE}.}
	\label{fig:pre_vs_rorr}
\end{figure*}
In this work, we benchmarked several approximate k-table mixing methods to the RORR method (see Sect.~\ref{sec:RORR}). As discussed in Sect.~\ref{sec:pre}, if the atmosphere is in chemical equilibrium, the abundances can be directly constrained as a function of pressure and temperature. It is then possible to use these abundances to create a grid of k-tables using the RORR method. This step can be done as a preprocessing step, so that the grid can be used as a lookup table to interpolate on during runtime. As mentioned in \citet{Amundsen2017ck}, the accuracy of this approach will be greatly dependent on the resolution of the grid. We match the pressure coordinates with our premixed tables in \texttt{expeRT/MITgcm}, removing the need for interpolation in pressure and thereby resulting in a higher accuracy and faster runtime. We then use 1000 temperature grid points in this work and in \citet{Schneider2022GCM} to further maximize the accuracy. 

In Fig.~\ref{fig:pre_vs_rorr} we show the temperature pressure profiles of the RORR simulation compared to the premixed (PRE) simulation. We find that both methods result in temperature profiles that overlap very well, with an error of less than 0.5\%. Premixed tables will therefore stay the method of choice for setups, where equilibrium chemistry can be assumed, given their much faster performance (see Table~\ref{tab:ctime}).

\FloatBarrier  
\section{A simple algorithm to calculate rainout}\label{app:rainout}
\begin{algorithm}
\caption{Procedure that takes in the pressure layers $p$ (sorted from bottom to top) and mass mixing ratios $\eta$ of an opacity species and returns the altered mass mixing ratios, under consideration of rainout.}
\label{alg:rainout}
\begin{algorithmic}
\Procedure{do\_rainout}{$p, \eta$}
    \State $\tilde\eta \gets \text{copy}(\eta)$
    \State $\text{species\_appearing} \gets \text{False}$
    \State $\text{species\_cond} \gets \text{False}$
    
    \For{$i \gets 2$ \textbf{to} $\text{length}(p)-1$} \Comment{Loop from bottom to top}
        \State $\text{grad} \gets -0.5 \times \left(\frac{{\eta_{i-1}-\eta_{i}}}{{p_{i-1}-p_{i}}} + \frac{{\eta_{i}-\eta_{i+1}}}{{p_{i}-p_{i+1}}}\right)$
        
        \If{$(\text{grad} > 0)$} \Comment{abundance increases}
            \State $\text{species\_appearing} \gets \text{True}$
        \EndIf
        
        \If{$(\text{species\_appearing})$} 
            \If{$(\text{grad} < 0)$} \Comment{abundance decreases again}
                \State $\text{species\_cond} \gets \text{True}$
            \EndIf
        \EndIf
        
        \If{$(\text{species\_cond})$} \Comment{Do the rainout}
            \If{$(\tilde\eta_{i-1} < \tilde\eta_{i})$}
                \State $\tilde\eta_{i} \gets \tilde\eta_{i-1}$
            \EndIf
        \EndIf
    \EndFor
    
    \If{$(\text{species\_cond})$} \Comment{Treat the boundary}
        \If{$(\tilde\eta_{-2} < \tilde\eta_{-1})$}
            \State $\tilde\eta_{-1} \gets \tilde\eta_{-2}$
        \EndIf
    \EndIf
    
    \State \textbf{return} $\tilde\eta$
\EndProcedure
\end{algorithmic}
\end{algorithm}

In this work, we consider the rainout of TiO and VO in order to demonstrate a possible use case of the DS method (Sect.~\ref{sec:rain}). We consider a species rained out, if, going upward from the bottom of the computational domain, the species has, in local chemical equilibrium, first become available in the gas phase and has then disappeared further up in the atmosphere. If the species became more abundant in the gas in local chemical equilibrium, we would not allow this and instead keep the fixed abundance from the layers below. These calculations are in practice performed by looking at the gradient of the mass mixing ratios, and we outline the algorithm used in this work in Algorithm~\ref{alg:rainout}. We would like to note that the algorithm presented here is likely oversimplified. Nevertheless, it provides mass mixing ratios that deviate from chemical equilibrium, and thus serves for demonstration purpose.

\FloatBarrier  
\section{Accuracy of fluxes and heating rates}\label{app:acc}
\begin{figure}
	\centering
	\includegraphics{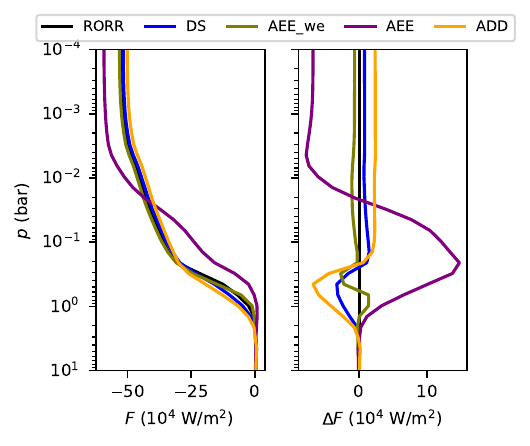}
	\caption{Total fluxes integrated over frequency at the substellar point (left) and difference to the RORR simulation (right).}
	\label{fig:flux_day}
\end{figure}
\begin{figure}
	\centering
	\includegraphics{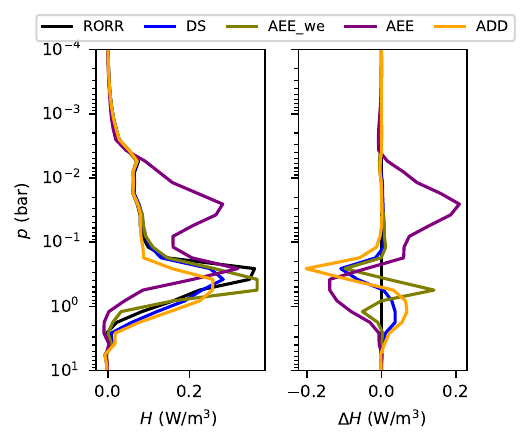}
	\caption{Heating rates (see Eq.~\ref{eq:hr}) at the substellar point (left) and differences to the RORR simulation (right).}
	\label{fig:hr_day}
\end{figure}
\begin{figure}
	\centering
	\includegraphics{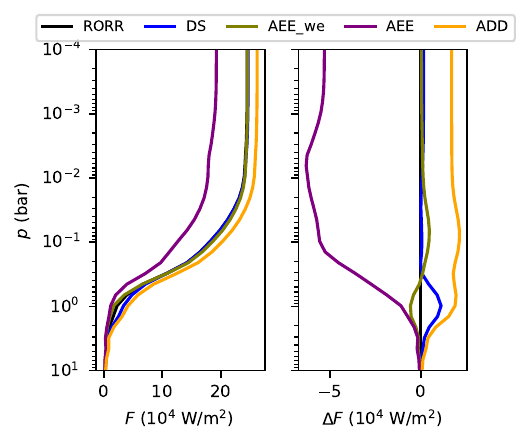}
	\caption{Total fluxes integrated over frequency at the antistellar point (left) and difference to the RORR simulation (right). The black dashed line in the right panel indicates the zero, which would mean no difference to RORR.}
	\label{fig:flux_night}
\end{figure}
\begin{figure}
	\centering
	\includegraphics{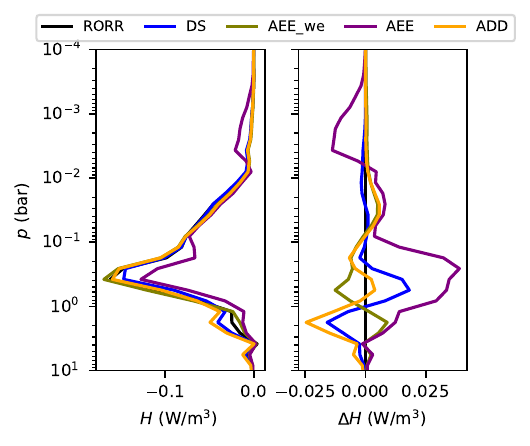}
	\caption{Heating rates (see Eq.~\ref{eq:hr}) at the antistellar point (left) and differences to the RORR simulation (right).}
	\label{fig:hr_night}	
\end{figure}

The thermal forcing in the GCM is given by the thermodynamic heating rate $H$, which is calculated as \citep[e.g.,][]{Amundsen2014GCM,Showman2009GCM}
\begin{equation}\label{eq:hr}
	H = -\frac{\mathrm{d}F}{\mathrm{d}z} = \frac{gp}{R_sT}\frac{\mathrm{d}F}{\mathrm{d}p},
\end{equation}
where $F$ is the net flux integrated over frequency, consisting of the bolometric stellar and planetary fluxes, $g$, $p$, $z$, and $R_s$ are the surface gravity, pressure, geometric height, and specific gas constant respectively. For a deeper understanding of the accuracy of the different approaches, we compare the accuracy of the resulting fluxes and heating rates of the individual methods in comparison to the RORR method. Since all the simulations lead to different atmospheric states, we opted to compare the fluxes and heating rates in a separate set of models. These models start from the final output of the RORR models at \SI{2000}{\day} and run for 10 radiative time-steps (a total of 1000 seconds). Unlike in the 2000-day-long simulations mentioned above, we did not limit the amount of scattering iterations in these extra runs. Using one radiative time-step for the comparison is not good enough, since the AEE\_we method needs the flux from the previous time-step. The advantage of this approach is that the temperature profiles are identical in these models. The comparison of fluxes and heating rates is thus fairer. We show the fluxes and heating rates for the substellar point in Figs.~\ref{fig:flux_day} and \ref{fig:hr_day} respectively, and those of the antistellar point in Figs.~\ref{fig:flux_night} and \ref{fig:hr_night} respectively. 

Looking at the day side (substellar point), we can see that AEE\_we and DS result in equally accurate heating rates and fluxes, followed by the ADD simulation, which also results in reasonable fluxes and heating rates. On the other hand, the non-weighted AEE simulations result in completely wrong fluxes and thus heating rates. The total flux at the substellar point is dominated by the stellar flux. We can see that the AEE method results in larger negative values of the stellar flux, again hinting to too little absorption in the upper parts of the atmosphere, as discussed in Sect.~\ref{sec:results}. Conversely, due to the overestimation of $\kappa_\mathrm{tot}$ at small $g$, almost all radiation is absorbed within a fine pressure range. While these trends can also be seen in the weighted AEE method, it seems to cause much less of an error, if the major absorber is wisely picked. 

Overall, we find that considering the substellar and antistellar fluxes and heating rates, the DS and AEE\_we method reproduce similarly accurate results. One of the reasons, why the DS method yields an overall better accuracy on the final atmospheric state might be found in the systematic error of the AEE method, which enhances $\kappa_\mathrm{tot}$ at small $g$ and decreases $\kappa_\mathrm{tot}$ at large $g$, which is not found in the DS method, which exhibits a random noise on the different $g$ values in each frequency bin (see Appendix~\ref{app:gval} for a discussion). These diagnostics reinforce that adaptive equivalent extinction should only be used together with a properly weighted average opacity. 
\end{appendix}
\end{document}